\documentclass[
reprint,
preprintnumbers,
nofootinbib,
bibnotes,
 amsmath,amssymb,
 aps,
 showpacs,
 superscriptaddress
pra,
floatfix,
]{revtex4-2}  

\usepackage{etoolbox}
\usepackage [T1]{fontenc}
\usepackage{bm}
\usepackage{color}
\usepackage{amsmath,amsfonts,amssymb,amscd,bm, mathrsfs}
\usepackage{graphicx}
\usepackage{eucal}
\usepackage{comment} 
\usepackage{enumitem}
\setlist[itemize]{noitemsep, topsep=0pt}   
\usepackage{placeins}  
\usepackage{appendix}
\usepackage{xcolor} 
\usepackage{graphicx}

\begin{document}
\newcommand{\Ci}{\textbf{[cite]}}
\newcommand{\Fi}{\textbf{[Add fig]}}

\title{Magnetic domain wall curvature effects on the velocity in the creep regime}
\author{Adriano Di Pietro}
\email{a.dipietro@inrim.it}
\author{Alessandro Magni}
\author{Gianfranco Durin}
\author{Michaela Kuepferling }
\affiliation{Istituto Nazionale di Ricerca Metrologica, Torino 10135, Italy}

\author{Giovanni Carlotti}
\author{Marco Madami}
\affiliation{Dipartimento di Fisica e Geologia, Universit{\`a} di Perugia, Perugia 06123, Italy}

\author{Cristopher H. Marrows}
\affiliation{School of Physics and Astronomy, University of Leeds, Leeds, LS2 9JT, UK}

\author{Silvia Tacchi}
\affiliation{CNR-IOM, Sede Secondaria di Perugia, c/o Dipartimento di Fisica e Geologia, Universit{\`a} di Perugia, I-06123 Perugia, Italy}

\author{Emily Darwin}
\affiliation{Magnetic \& Functional Thin Films Laboratory, Empa,
	Swiss Federal Laboratories for Materials Science and Technology, 8600 D{\"u}bendorf, Switzerland}

\author{Alexandra J. Huxtable}
\affiliation{School of Physics and Astronomy, University of Leeds, Leeds, LS2 9JT, UK}

\author{Bryan J. Hickey}
\affiliation{School of Physics and Astronomy, University of Leeds, Leeds, LS2 9JT, UK}

\date{\today}

\begin{abstract}
We study the effect of magnetic domain wall curvature on its dynamics in the creep regime in systems displaying Dzyaloshinskii-Moriya interaction (DMI). We first derive an extended creep model able to account for the finite curvature effect in magnetic bubble domain expansion. We then discuss the relative importance of this effect and discuss its dependence on the main magnetic and disorder parameters. We show this effect can be easily measured in Pt/Co multi layer samples, reporting a strong velocity reduction below a sample dependent threshold value of the magnetic bubble radius. We finally show, both theoretically and experimentally, how the radius of magnetic bubbles can have a strong impact on the reproducibility of DMI measurements. 
\end{abstract}

\maketitle

\section{Introduction}
\label{Sec:Introduction}
Magnetic domain wall (DW) motion in ferromagnetic thin films is a complex phenomenon displaying a variety of dynamical regimes, identified as creep, depinning, steady flow and precessional flow \cite{Metaxas2007,Cayssol2004,herrera2015controlling,ferrero2021creep}. Depending on which regime we are accessing, the measurement of the DW velocity in response to externally applied fields allows to extract some key magnetic parameters of the sample. Among these magnetic parameters is the Dzyaloshinskii-Moriya interaction (DMI) \cite{DZYALOSHINSKY1958241,moriya1960anisotropic} strength, which is becoming increasingly relevant because it can enhance the potential functionality of magnetic materials: we mention the stabilization of skyrmions \cite{fert2017magnetic,fert2013skyrmions,finocchio2016magnetic,bogdanov1994thermodynamically} as potential information carriers for in memory computing platforms as well as enhanced domain wall (DW) velocity, useful to increase the performances of magnetic racetrack memory devices \cite{thiaville_dynamics_2012}. Measuring the DW velocity in the creep regime has the advantage of being possible under very easily accessible experimental conditions (low fields, long pulse duration, room temperature), but has the downside of a general lack of generality of the models describing the DW dynamics \cite{MAG2022,Kuepferling2023,Vaatka2015,Hartmann2019}. The measurement of DMI strength via DW velocity methods in the creep regime \cite{JE2013,Pakam2024,Hartmann2019} is a clear example of the need for more accurate models, as there have been accounts of large discrepancies in the reported DMI strength values, especially when comparing these values with those obtained via Brillouin light scattering (BLS) \cite{Kuepferling2023} or other disorder independent methods (such as measurements in the flow regime) \cite{Garcia2021}.
One factor that has been seldom inspected in the modelization of DW dynamics in the creep regime curvature of the DW \cite{MOO2011,zhang_domain-wall_2018,zhang_direct_2018}. In this paper, we address this issue by investigating the DW dynamics of magnetic bubble domains, taking into account the initial radius of the nucleated bubble. We demonstrate that the size of the circular magnetic domain in the creep regime can play a significant role in determining the speed of the magnetic domain wall. As a consequence, all the associated properties that are measured on the basis of the creep velocity require a precise knowledge of the initial bubble radius. In particular, we show how the consideration of a finite radius of the magnetic bubble can have very important consequences on reproducibility of measurements of the DMI strength.\\
The paper is organized as follows: In Section \ref{Sec:Theoretical_background}, we introduce the theoretical models describing magnetic bubbles and provide a brief overview of the theory of magnetic DW motion in the thermally activated (creep) regime. In \ref{subsec:Extenden_creep_model} we propose an extension of the creep model, which accounts for the DW curvature radius within the rigid bubble approximation. In Section \ref{sec:model-predictions} we report the model predictions, discussing which parameters are expected to have a greater impact on the relevance of the effect. In Section \ref{Sec:Materials_and_methods} we provide experimental details regarding the measurement techniques, the sample characteristics and the data analysis methods used to extract the DW velocities. In Section \ref{Sec:Results} we present our experimental results and show how our model can interpret the experimentally measured decreased DW velocity for small bubbles. We subsequently demonstrate how the incorporation of the bubble radius in the DW velocity model can improve the reproducibility of DMI measurements when dealing with bubbles with variable sizes. Finally, in Section \ref{sec:conclusions}, we provide our conclusions and an outlook on future explorations regarding the radius dependence of the DW velocity.

\section{Theoretical background}
\label{Sec:Theoretical_background}
\begin{figure}[t]   
 \centering
    \includegraphics[width=\linewidth]{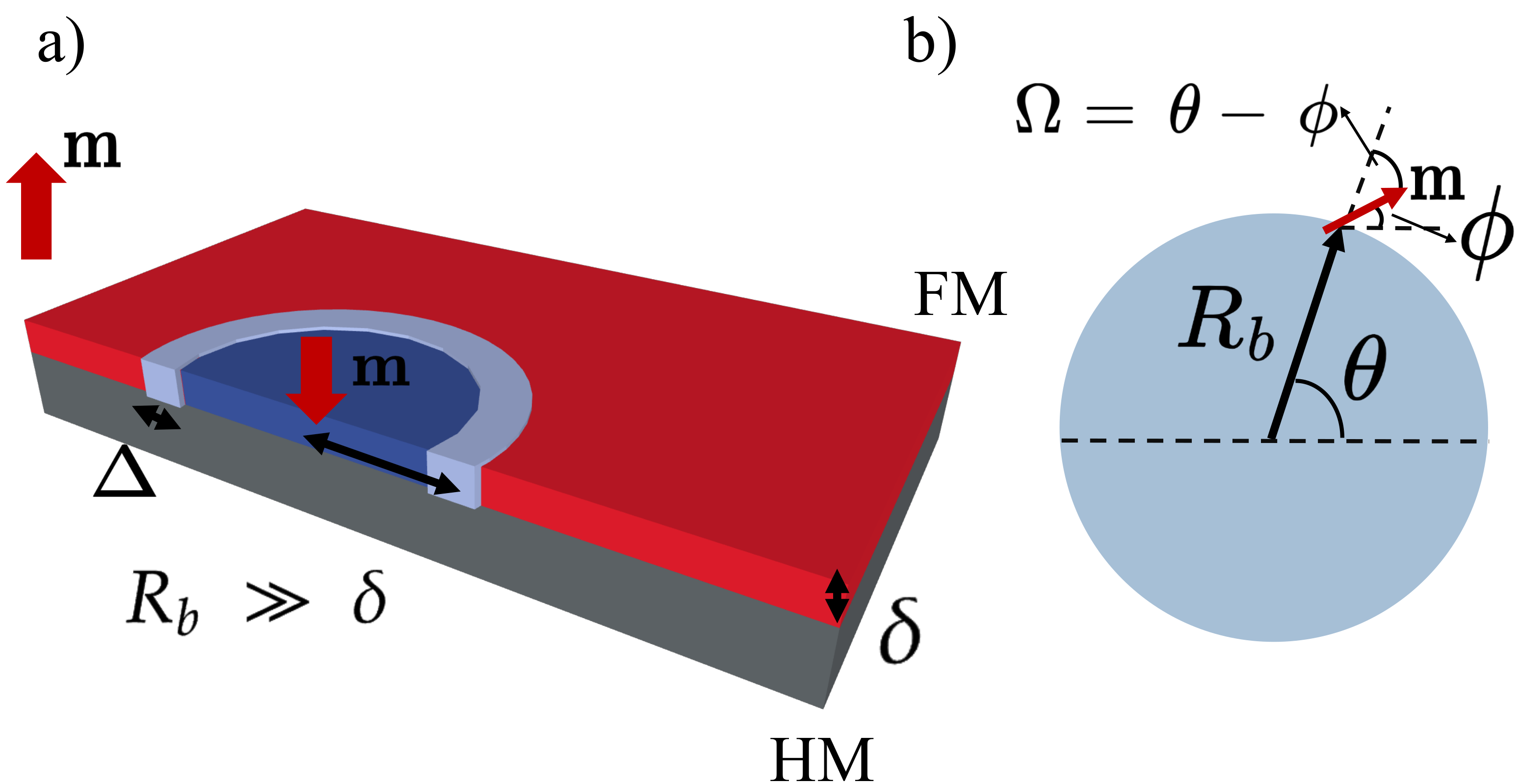}
    \caption{ a) Bubble domain in a typical bilayer structure (heavy metal - HM / Ferromagnet -  FM) with DMI and PMA. $\Delta$ Represents the DW width, $R_b$ represents the bubble radius and $\delta \ll R_b$, the sample thickness. The red and blue regions represent areas of opposite magnetization. b) Magnetic bubble parameters: $\theta$ represents the angle that parametrizes the location of the bubble edge and $\phi$ represents the IP magnetization angle in the middle of the DW, i.e. where we do not have a vertical component $m_z$. } 
    \label{fig:Setup_bubble}
\end{figure}

\begin{figure}
    \centering
    \includegraphics[width=0.8\linewidth]{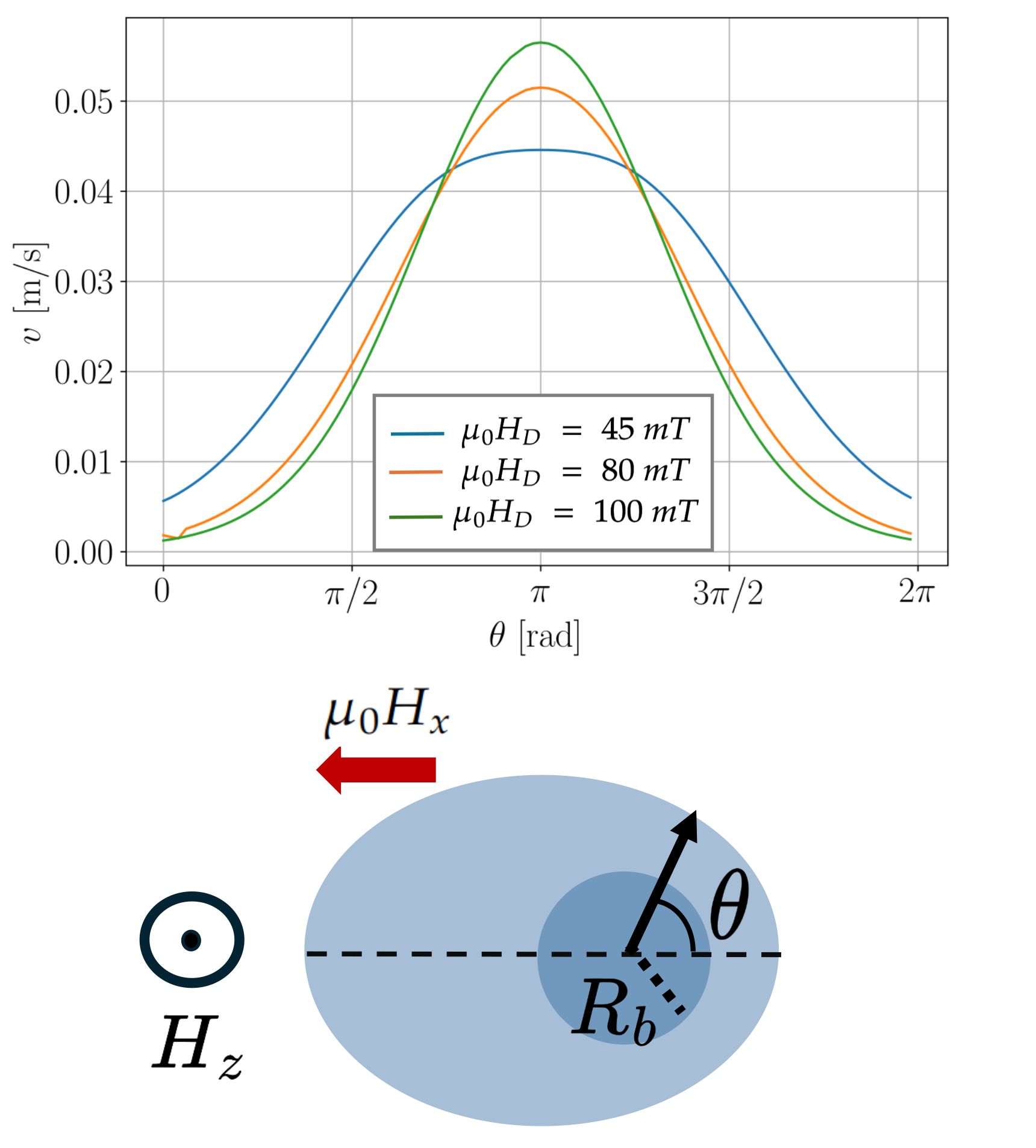}
    \caption{(Top) DW velocity as a function of the angle $\theta$ (see Eq.\eqref{eq:DW_1D}) for 3 different values of the DMI field with a constant applied IP field $\mu_0 H_x=$ -80 mT. (Bottom) Cartoon representing the asymmetric bubble expansion in the presence of DMI and IP applied field $\mu_0 H_x$ and an OOP field $\mu_0 H_z$. The parameters used are $M_s = 1.2 $ MA/m , $K_{\text{eff}} = 0.513$ MJ/m$^3$, $\alpha_0 = 15.47$ T$^{1/4}$, $\ln v_0 = 40.5$. and the OOP field is given by $\mu_0 H_z = 13.4$ mT. $R_b$ represents the radius of the initially nucleated bubble. }
    \label{fig:DW_equilibrium_and_velocity}
\end{figure}

As mentioned in the Introduction  \ref{Sec:Introduction}, the dynamics of a magnetic DW under an applied out-of-plane (OOP) field displays several distinct regimes \cite{LEM1998,chauve2000creep}. While the flow regime (and beyond) can effectively be described without taking in account the disorder potential \cite{moore2008high,jue2016domain} in which the magnetic domain wall moves, in the creep regime the DW follows an Arrhenius-type law \cite{blatter1994vortices} $v \propto \exp(- F_b / k_B T)$, where the velocity is determined by the competition 
between the energy barrier $F_b$ imposed by the pinning potential and the thermal energy $k_B T$ (see Appendix \ref{Appendix B: : Brief reminder of creep theory}). When a driving force $f$ such as an applied OOP magnetic field is applied, the energy barrier scales with the creep power law $F_b \propto f^{-\mu}$, where $\mu$ is the creep critical exponent that depends on fundamental properties of the system such as its dimensionality and the range of interactions \cite{LEM1998}. In typical ferromagnetic thin films, the value of the creep critical exponent is $\mu = 1/4$ \cite{barabasi1995fractal,ji1991transition}.  
The study of domain wall dynamics in the creep regime is crucial as it can reveal key magnetic properties under easily accessible experimental conditions. One particularly important property that can be extracted from the DW velocity is the DMI strength. This, among other methods, can be achieved by utilizing the symmetry-breaking effects of an additional in-plane (IP) magnetic field, which, in the presence of DMI, induces a measurable distortion of the magnetic bubble \cite{JE2013}. This asymmetric expansion along the profile of the bubble originates from the modification of the domain wall energy density caused by the interplay between DMI and the applied IP field \cite{thiaville_dynamics_2012,Pakam2024}. Assuming a magnetic bubble domain in the presence of perpendicular magnetic anisotropy (PMA), interfacial Dzyaloshinskii-Moriya interaction and applied IP field along the bubble profile, the DW surface energy density $\bar{\sigma}_{DW}$ ($[\bar{\sigma}_{DW}]$ = J/m$^2$) is given by \cite{Pakam2024} 
\begin{align}
    \bar{\sigma}_{DW}(H_x,\theta, \phi) &=  4 \sqrt{A K_{\text{eff}}} - \pi M_s \Delta \mu_0(H_D \cos(\theta - \phi) \nonumber \\ &+  H_x\cos(\phi)) +\frac{\ln(2)}{\pi} \delta \mu_0 M_s^2 \cos^2(\theta - \phi) ,
\label{eq:DW_1D}\end{align}
where $A$ represents exchange stiffness, $K_{\text{eff}}$ the effective anisotropy, $M_s$ the saturation magnetization, $\Delta = \sqrt{A/K_{\text{eff}}}$ the DW width, $H_D$  the DMI effective field, $\delta$ the sample thickness and $ \{\phi, \theta \}$ represent respectively, the domain wall magnetization angle and the angular coordinate along the circumference of the bubble (see Fig.\ref{fig:Setup_bubble}-b).
We identify the angle $\theta$ dependent equilibrium energy density of the magnetic bubble as 
\begin{align}
    \bar{\mathcal{E}}_{DW}(H_x,\theta) = \min_{\phi \in [0,2 \pi)} \bar{\sigma}_{DW}(H_x,\theta, \phi). 
\end{align}
This equilibrium domain wall energy density translates to the angle dependent velocity $v(\theta)$ of the bubble according to \cite{pakam_anisotropic_2024}
\begin{align}
    v(\theta, H_z) =v_0 \exp\bigg( - \alpha_0 \bigg(\frac{\bar{\mathcal{E}}_{DW}(H_x,\theta)}{\bar{\mathcal{E}}_{DW}(0,\theta)\mu_0 H_z} \bigg)^{1/4}\bigg),
\label{eq:v_Hz_standard}\end{align}
where $v_0$ and $\alpha_0$ are experimentally measured creep parameters (see \ref{Sec:Materials_and_methods}). $v(\theta)$ can then be used to fit the experimentally determined angle dependent velocity of the DW and extract important physical parameters of the sample such as the DMI strength  \cite{Pakam2024} (see Fig.\ref{fig:DW_equilibrium_and_velocity}). 
\subsection{Extended creep model}
\label{subsec:Extenden_creep_model}
\begin{figure*}
    \centering
    \includegraphics[width=0.75\linewidth]{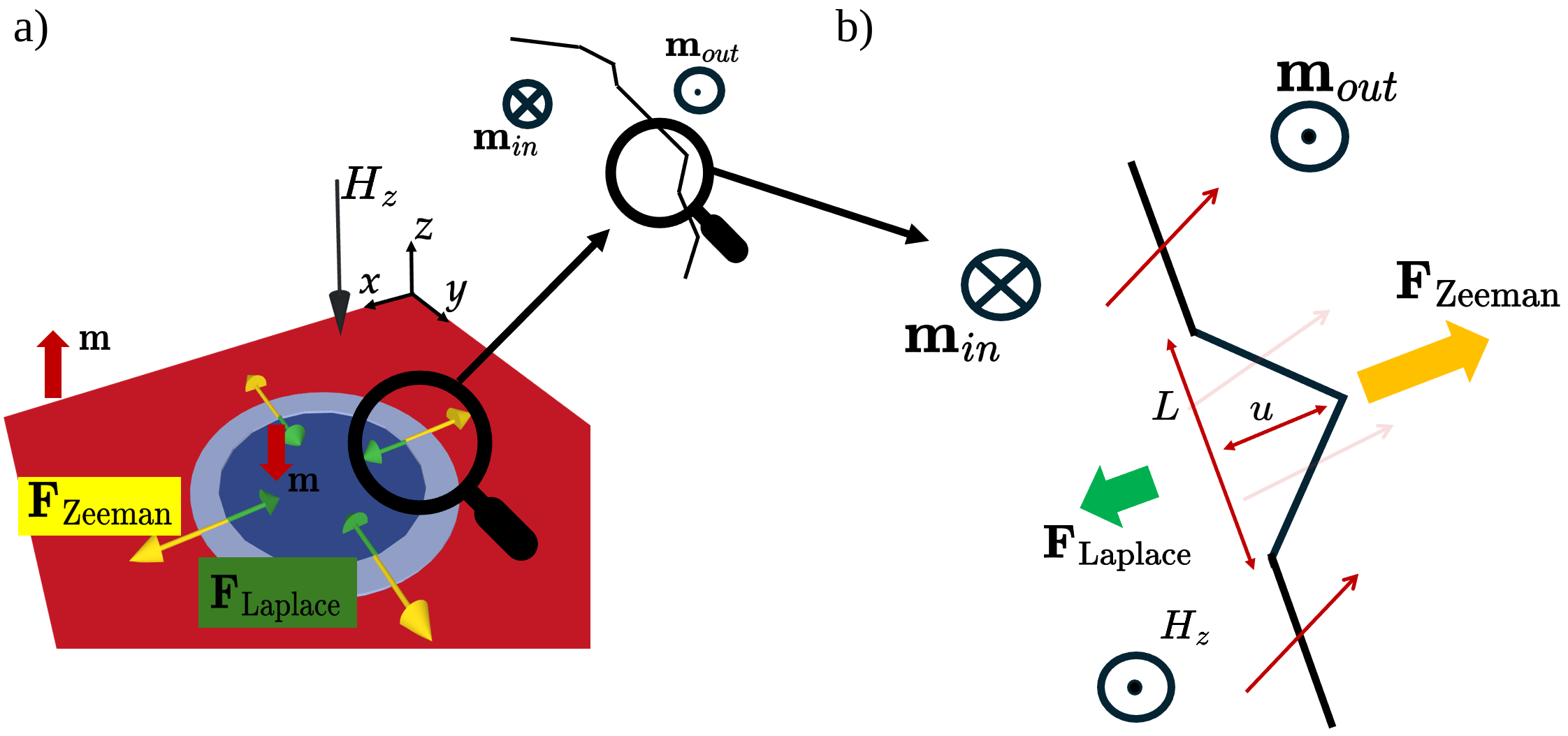}
    \caption{a) Macroscopic picture of the force balance at the boundary of a magnetic bubble in the presence of an applied magnetic field H$_z$ and a competing Laplace pressure generated by the surface tension of the bubble. The inset depicts a more realistic picture of an actual boundary between domains \cite{LEM1998,Hartmann2019} (observe the irregular pattern) b) Microscopic picture of the DW segment deformation in response to the application of a driving force ($\bm{F}_{Zeeman}$) and a competing Laplace force ($\bm{F}_{Laplace}$). $L$ and $u$ represent the geometrical parameters of the DW deformation. }
    \label{fig:Laplace_balance}
\end{figure*}

Up to this point, we have completely omitted any radius dependence in the energy density, effectively treating the domain wall velocity with a true 1D model \cite{JE2013}. If instead we introduce a domain wall energy density that takes in account the bubble curvature \cite{VANDERMEULEN2018337, MorettiPhD} (see Appendix \ref{Appendix A:Derivation of domain wall energy density for a bubble domain} for a brief reminder of the derivation),the energy density of Eq.\eqref{eq:DW_1D} becomes
\begin{align}
    &\sigma_{DW}(H_x,\theta,\phi, R_b) = \bigg[4\sqrt{A K_{\text{eff}}}- \pi M_s \Delta \mu_0(H_{D} \cos(\theta - \phi) \nonumber \\ 
    &+  H_x\cos(\phi)) + \frac{\ln(2)}{\pi} \delta \mu_0 M_s^2 \cos^2(\theta - \phi) \bigg] + \frac{2 A \Delta}{R_b^2} \\ &= \bar{\sigma}_{DW} + \frac{2 A \Delta}{R_b^2},
\end{align}
where we have defined $\bar{\sigma}_{DW}$ as the radius independent part of the energy density, equivalent to Eq.\eqref{eq:DW_1D}. The corresponding equilibrium energy density of the DW at each angle $\theta$ is
\begin{align}
    \mathcal E_{DW}(\theta, H_x, R_b) &=  \min_{\phi \in [0, 2 \pi)} \sigma_{DW}(H_x,\theta,\phi, R_b) \nonumber \\  &= \bar{\mathcal{E}}_{DW} + \frac{2 A \Delta}{R_b^2}.  \label{eq:Equilibrium_E_DW}
\end{align}
We can now write the equations of motion of a DW parameterized by the collective coordinates $\{R_b(t),\Omega(t)\}$ by expressing the Lagrangian density $\mathcal L$ and the Rayleigh dissipation function $\mathcal F$ as follows \cite{Ciornei2011,ThiavilleBook2006}
\begin{align} \mathcal L &= \sigma_{DW} + \frac{\mu_0 M_s}{\gamma_0} 2 \pi \dot \Omega R_b^2 , \\\mathcal F &= \frac{2 \pi \alpha \mu_0 M_s R_b}{\gamma_0}\bigg( \frac{\dot{R}_b^2}{\Delta} + \Delta \dot{\Omega}^2\bigg). \end{align}
Since we are primarily interested in the dynamics of the DW coordinate $R_b(t)$ (i.e. the velocity of the DW), we neglect the dynamics of the magnetization angle $\Omega(t)$ and only derive the DW velocity $\dot R_b$ using the Euler-Lagrange-Rayleigh equation
\begin{align}&\dot R_b \propto \alpha \bigg[ H_z - \frac{1}{2 \mu_0 M_s R_b}\bigg(\bar{\sigma}_{DW} - \frac{2 A \Delta}{R_b^2}\bigg)\bigg] \nonumber \\ &+ \frac{1}{2} \bigg(\pi  H_{D} \sin \Omega - \frac{\ln(2)}{2 \pi \Delta} \delta M_s \sin(2 \Omega) \bigg).  \label{eq:Rb_dot}\end{align} 
Neglecting higher orders in $1/R_b$, we can rewrite the equation of the DW velocity as 
\begin{align}
    &\dot R_b \propto \alpha \bigg[ H_z - \frac{\bar{\sigma}_{DW}}{2 \mu_0 M_s R_b} \bigg] \nonumber \\ &+ \frac{1}{2} \bigg(\pi H_{D} \sin \Omega - \frac{\ln(2)}{2 \pi \Delta} \delta M_s \sin(2 \Omega) \bigg) + \mathcal{O}\bigg( \frac{1}{R^2_b} \bigg) \label{eq:modified_velocity_Rb}
\end{align}
At this point we can observe how the term $\frac{\bar{\sigma}_{DW}}{2\mu_0 M_s R_b}$ can be interpreted as a Laplace pressure term \cite{zhang_direct_2018,zhang_domain-wall_2018} competing with the external field $H_z$ (see Fig.\ref{fig:Laplace_balance}). In the following, we show what effects this additional term might have on the velocity of the DW in the creep regime, and the relative consequences for the determination of physical quantities related to it. First of all, let us consider a straight DW which is deformed through the application of an external force such as a magnetic field applied $H_z$ perpendicular to the plane of the sample. We shall assume that the lateral deformation is of size $L$ and the longitudinal displacement of the wall is identified by $u$ (see Fig.\ref{fig:Laplace_balance}). The energy balance $\Delta E_{balance} (u,L)$ determining the dynamics of the DW in the creep regime is expressed by the competition of:
\begin{itemize}
    \item Zeeman energy gain: $$- \mu_0 M_s H_z \frac{\delta L u}{2}$$
    \item Elastic energy cost: $$ 2 \frac{u^2}{\delta L} \mathcal{E}_{DW} $$
    \item Pinning energy gain: $$- \sqrt{\gamma \xi^2 L},$$
\end{itemize}
where $\gamma$ has units $[\gamma] = J^2/m^3$ and represents the typical pinning potential strength, while $\xi$ has units $[\xi] = m$ and represents the correlation length of the pinning centers \cite{LEM1998,chauve2000creep,Hartmann2019}.

If  we now consider the finite radius of the bubble, according to Eq.\eqref{eq:modified_velocity_Rb}, there is an additional term acting on the energy balance in a fashion  similar to the hydro-static pressure (or Laplace pressure) acting on the surface of a soap bubble \cite{zhang_direct_2018,zhang_domain-wall_2018,MOO2011} 
\begin{align}
   \frac{\bar{\mathcal{E}}_{DW} \delta L u}{2 R_b},
\end{align} 
where $R_b$ represents the radius of the magnetic bubble. As can be seen by the above expression, the Laplace pressure term competes with the driving force $\propto H_z$ and can therefore be included in the energy balance in the straightforward fashion 
\begin{align}
    &\Delta E_{balance} (u ,L) = \nonumber \\  & \mathcal{E}_{DW} \frac{2 u^2}{\delta L} - \sqrt{\gamma \xi^2 L} + \bigg(\frac{\bar{\mathcal{E}}_{DW}}{R_b} - M_s H_z \mu_0 \bigg)\frac{\delta L u}{2},
\end{align}
which can easily be recast in the form \cite{Hartmann2019}
\begin{align}
    &\Delta E_{balance} (u ,L) = \nonumber \\ &\mathcal{E}_{DW} \frac{2 u^2}{\delta L} - \sqrt{\gamma \xi^2 L} -\mu_0 M_s \tilde{H}_z(R_b) \frac{\delta L u}{2}, 
    \label{eq:Energy_barrier}
\end{align}
by modifying the applied field $H_z$ by an $R_b$ dependent effective field $\tilde{H}_z(R_b)$
\begin{equation}
    \tilde{H}_z(R_b) =  \bigg(H_z - \frac{\bar{\mathcal{E}}_{DW}}{\mu_0 M_s R_b} \bigg).
\end{equation}
The advantage of this approach is that it allows to generalize the creep theory immediately by modifying the force term $H_z^{-1/4} \rightarrow \tilde{H}_z(R_b)^{-1/4} $  in the expression of the velocity in the creep regime of Eq.\eqref{eq:v_Hz_standard}. The DW velocity as a function of the initial radius $R_b$ of the bubble domain can be expressed as 
\begin{align}
    &v(\theta,H_x,H_z,R_b) = \nonumber\\ &v_0 \exp \bigg[ - \alpha_0 \bigg(\frac{(\bar{\mathcal{E}}_{DW}(\theta,H_x) + 2A\Delta/R_b^2)}{\tilde{H}_z(R_b)(\bar{\mathcal{E}}_{DW}(\theta,0) + 2A\Delta/R_b^2 ) } \bigg)^{1/4} \bigg],
    \label{eq:DW_Rb}
\end{align}

where we have used the fact $\mathcal{E}_{DW} = \bar{\mathcal{E}}_{DW}+  2A\Delta/R_b^2$ to make the radius dependence of the velocity explicit. To compare the velocity behavior across a variety of different samples, it is useful to normalize Eq.\eqref{eq:DW_Rb} with respect to the value of the velocity in the case of a straight DW, i.e. an magnetic bubble with an approximately infinite radius. In the absence of applied IP fields we can write 
\begin{align}
    &\frac{v(0,0,H_z,R_b)}{v(0,0,H_z,R_b\rightarrow \infty)} = \nonumber \\  &\exp \bigg[ - \alpha_0 ( \tilde{H}_z(R_b)^{-1/4} - H_z^{-1/4} )  \bigg]. \label{eq:v_Rb}
\end{align}
To assess the effect of the different physical parameters on the strength of the radius dependent effect, we can invert Eq.\eqref{eq:v_Rb} and define $\tilde{R}_b$ as the radius value for which the DW velocity is halved, i.e. 
\begin{align}
    \frac{v(\tilde{R}_b)}{v(R_b \rightarrow \infty)} &:= \frac{1}{2} \\ \Rightarrow \tilde{R}_b &= \frac{\bar{\mathcal{E}}_{DW}}{2 \mu_0 M_s} \frac{1}{\bigg[ H_z^{-1/4} + \frac{\ln{2}}{\alpha_0} \bigg]^{4}}. \label{eq:Rb_tilde}
\end{align} 
$\tilde{R}_b$ can function as a metric for the strength of the pressure effect on the dynamics of the DW. 
\section{Model predictions} \label{sec:model-predictions}
\begin{figure*}
	\centering
	\includegraphics[width=0.6\linewidth]{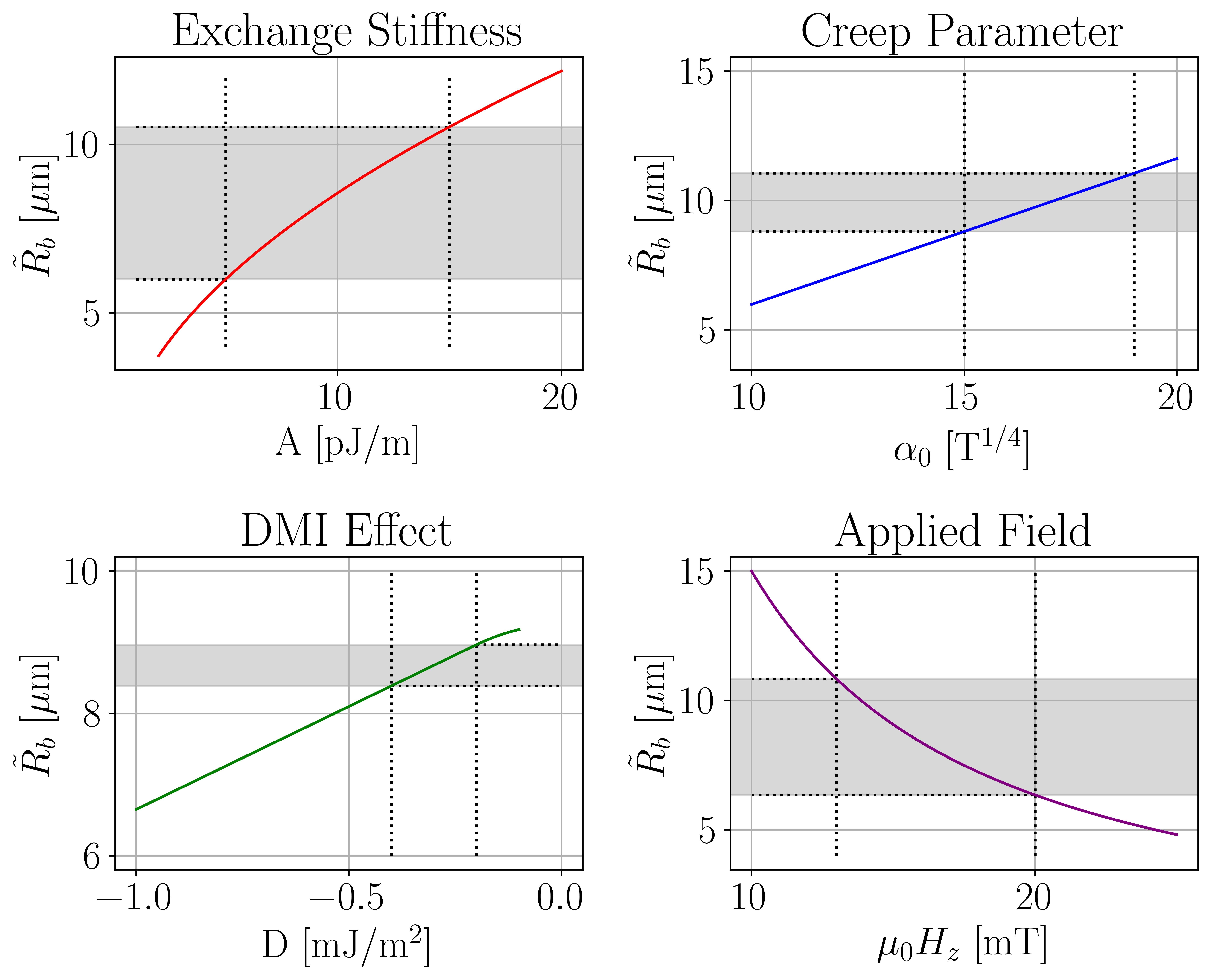}
	\caption{Effect of different physical parameters on $\tilde{R}_b$ of Eq.\eqref{eq:Rb_tilde}. Top left: Effect of exchange stiffness $A$, top Right: Effect of the creep parameter $\alpha_0$, bottom left: effect of DMI strength $D$, bottom right: applied field $\mu_0 H_z$. The applied OOP field in all cases except in the last subplot (bottom right) is $\mu_0 H_z = 15$ mT. The areas highlighted in gray are a guide to the estimated strength of the effect and refer the regions of interest (reported on the $x$-axis of the plots) for the experimental parameters of the present samples \cite{MAG2022}. The non-changing parameters used in the respective plots are $A = 11.2$ pJ/m, $M_s = 1.2 $ MA/m , $K_{\text{eff}} = 0.513$ MJ/m$^3$, $\alpha_0 = 15.47$ T$^{1/4}$, $\mu_0 H_{D} =-30$ mT. } 
	\label{fig:Rb_tilde}
\end{figure*}

As can be seen from Eq.\eqref{eq:Rb_tilde}, the strength of the Laplace pressure effect depends both on magnetic and disorder parameters. As a first step to understand the importance of the Laplace pressure term (see Eq.\eqref{eq:v_Rb}) for different samples with varying magnetic and disorder properties, we observe how the different physical parameters affect its strength. 
In Fig.\ref{fig:Rb_tilde} we show how exchange stiffness $A$, DMI strength $D$, creep parameter $\alpha_0$ and the applied field $\mu_0 H_z$ can affect the parameter $\tilde{R}_b$ (see Eq.\eqref{eq:Rb_tilde}).  

\begin{figure}
	\centering
	\includegraphics[width=1\linewidth]{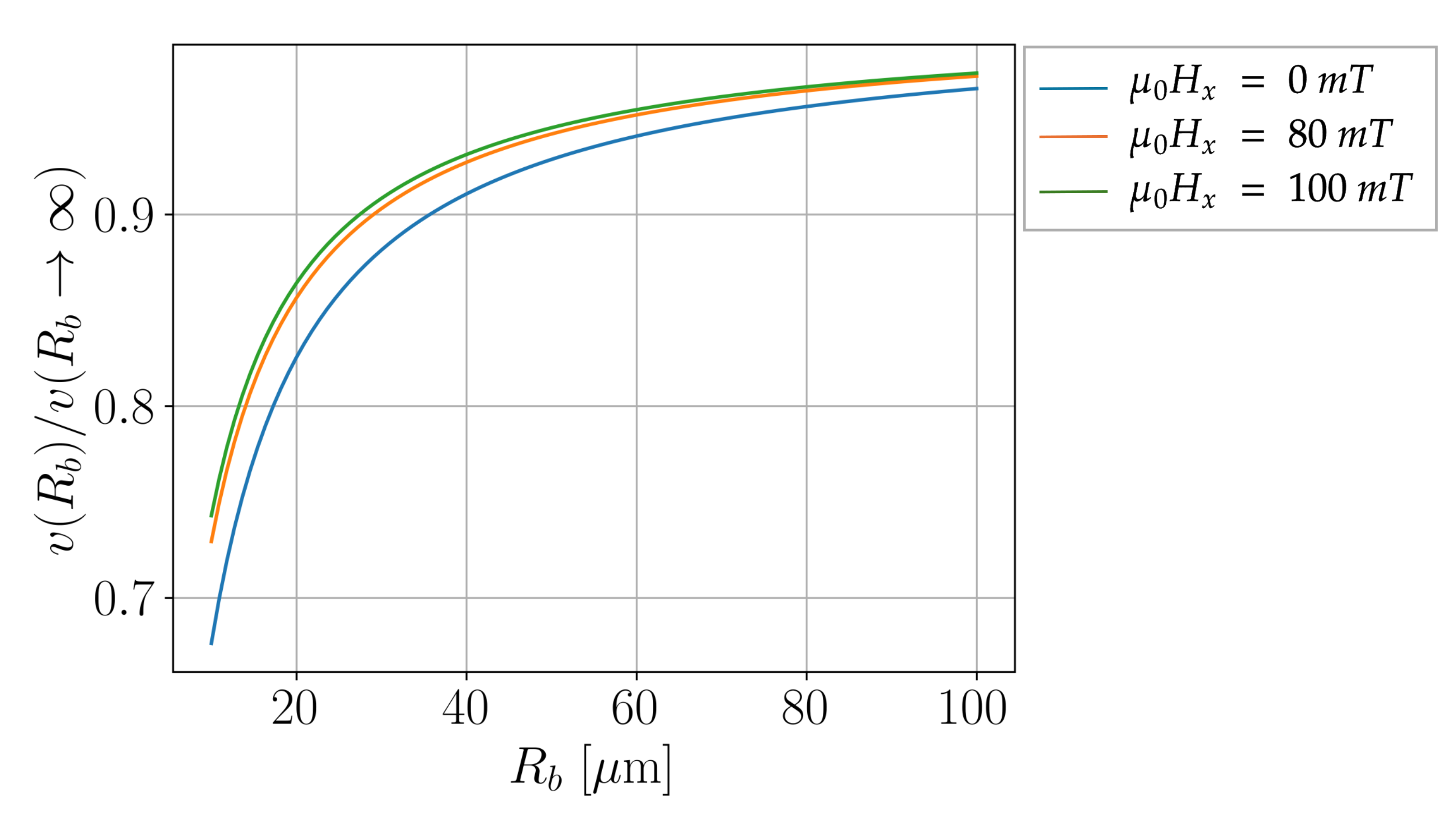}
	\caption{Normalized velocity $v(R_b)/v(R_b \rightarrow \infty)$ as a function of the initial bubble radius $R_b$ in the presence of an additional IP field $\mu_0 H_x$ as predicted by Eq.\eqref{eq:v_Rb}. The parameters used are $M_s = 1.2 $ MA/m , $K_{\text{eff}} = 0.513$ MJ/m$^3$, $\alpha_0 = 15.47$ T$^{1/4}$, $\ln v_0 = 40.5$, $\mu_0 H_D = -20$ mT, $\mu_0 H_z = 15.45$ mT. } 
	\label{fig:radiusdependencehx}
\end{figure}

\begin{figure}
	\centering
	\includegraphics[width=\linewidth]{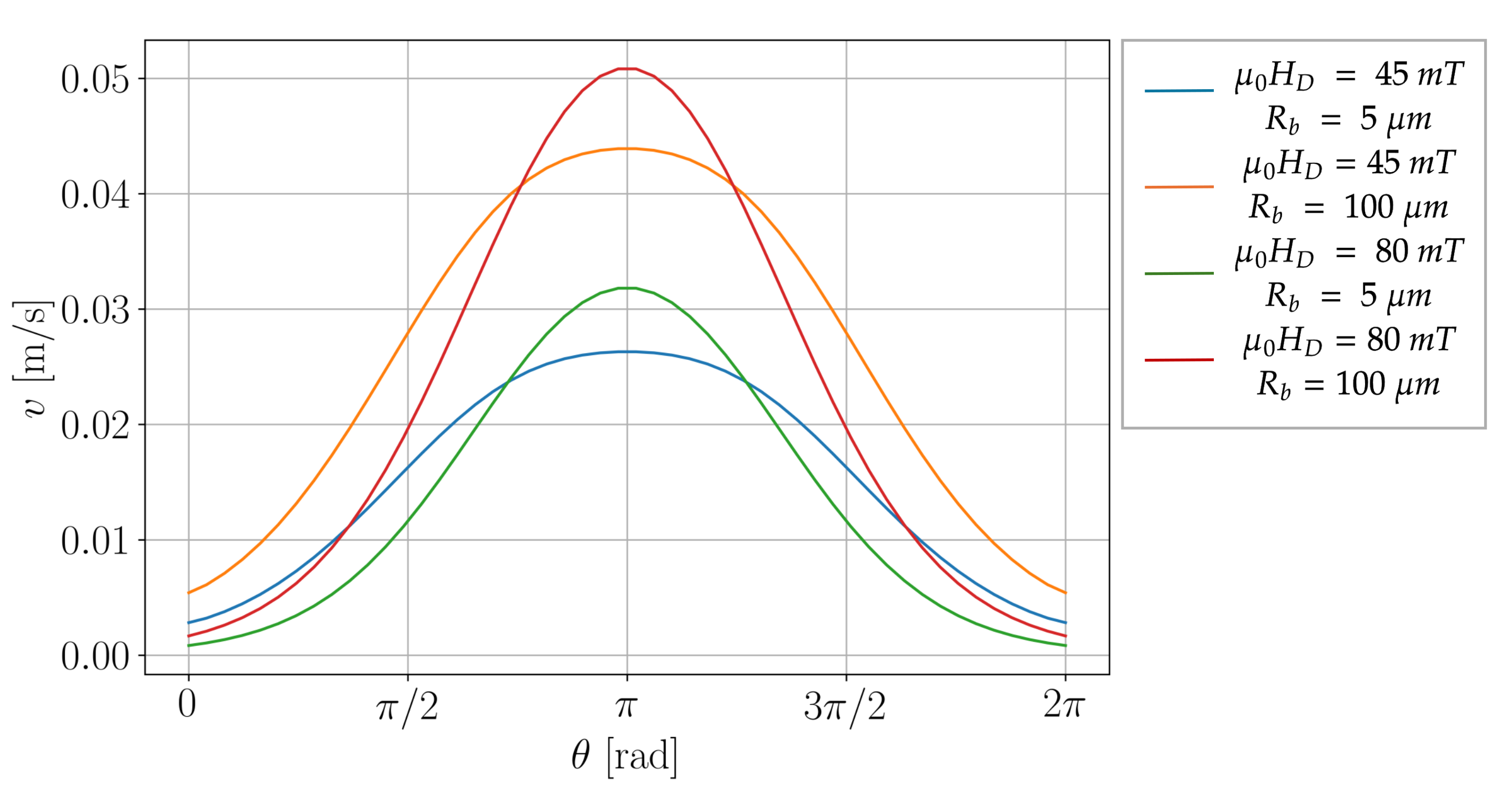}
	\caption{Effect of the finite radius on the velocity profile of the bubble according to Eq.\eqref{eq:DW_Rb} for different values of the DMI strength.The parameters used are $M_s = 1.2 $ MA/m , $K_{\text{eff}} = 0.513$ MJ/m$^3$, $\alpha_0 = 15.47$ T$^{1/4}$, $\ln v_0 = 40.5$, while the OOP field is $\mu_0 H_z = 13.4$ mT.}
	\label{fig:Rb_effect_on_V}
\end{figure}

\noindent The gray region in Fig.\ref{fig:Rb_tilde} indicates a typical range of physical parameters which might be observed in samples exhibiting bubble domains, such as the here studied Pt/Co multilayers. As we can see from Fig.\ref{fig:Rb_tilde}, the DMI strength has a smaller effect on the DW velocity when compared to the effect of the exchange stiffness $A$. We also emphasize how the creep parameter $\alpha_0$, which, at fixed $\bar{\mathcal{E}}_{DW}$, depends on a combination of the disorder parameters introduced in \ref{Sec:Theoretical_background} \cite{LEM1998,Hartmann2019} (see Appendix \ref{subsec:Bubble_expansions_method}), has a notable effect on $\tilde{R}_b$, showing how the radius dependence of the DW velocity in the creep regime can also be influenced by the disorder characteristics of the material and not only by its magnetic properties. \\
The determination of the DMI strength from the experimental measurement of DW velocity in the creep regime requires the concurrent presence of an IP field $H_x$, as this is necessary to introduce the asymmetry in the bubble expansion \cite{JE2013,Pakam2024}. The presence of this additional $H_x$ field indeed alters the strength of the effect on the radius dependence (see Fig.\ref{fig:radiusdependencehx}) and induces a noticeable change on the velocity profile of the bubble. In Fig.\ref{fig:Rb_effect_on_V} we can clearly see how the model predicts a marked change in the velocity profile of DW along the circumference of the bubble with different $\mu_0 H_{D}$ and initial radii $R_b$ combinations. Therefore the radius dependence cannot be neglected in models aiming at describing bubble DW motion in a more general sense (i.e. across different bubble sizes). 

\section{Experimental}
\label{Sec:Materials_and_methods}
\subsection{Materials and methods}
\begin{figure}[h]
	\centering
	\begin{minipage}{0.45\textwidth}
		\centering
		\includegraphics[width=\textwidth]{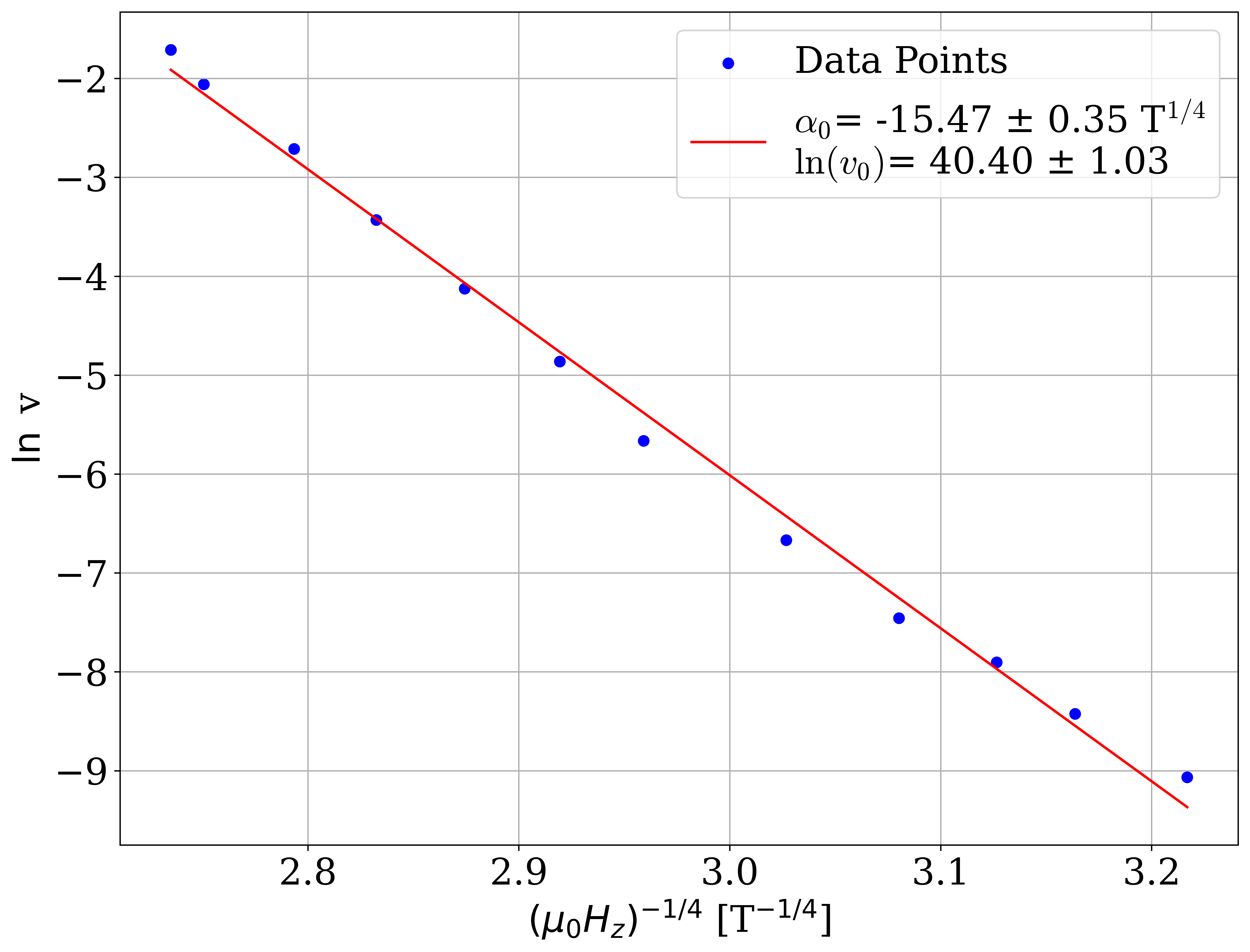}
		\label{fig:plot1}
	\end{minipage}
	\hfill
	\begin{minipage}{0.45\textwidth}
		\centering
		\includegraphics[width=\textwidth]{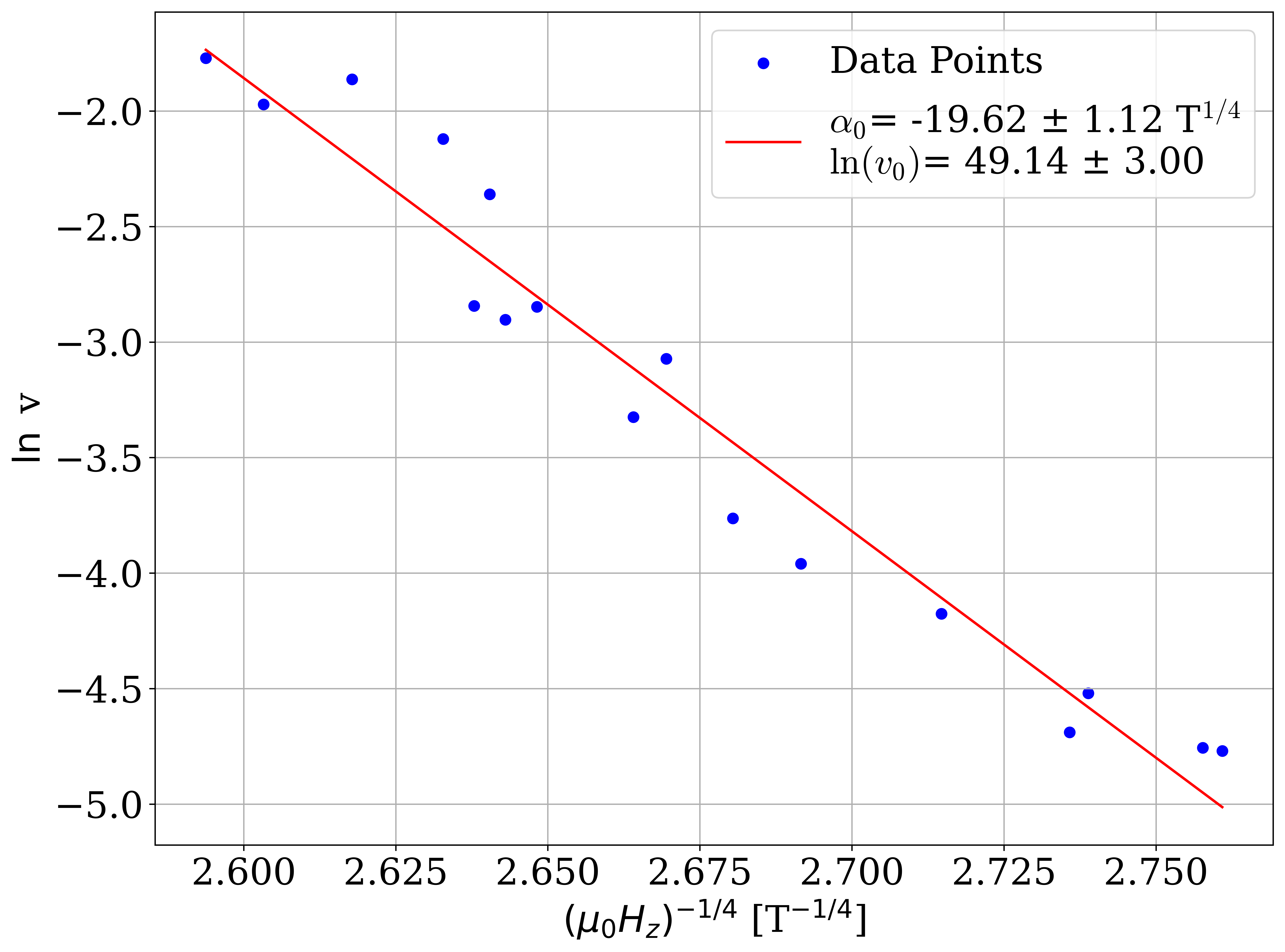}
		\label{fig:plot2}
	\end{minipage}
	\caption{Linear fit of Eq.\eqref{eq:creep_law_fits} on the experimentally measured values of the DW velocity $v$ as a function of the applied OOP field $\mu_0 H_z$ (see Eq.\eqref{eq:creep_law_fits}).  The top figure refers to the creep measurement performed on sample $a_4$ and the bottom figure represents the creep measurement performed on sample $a_3$ (see TABLE \ref{tab:sample_vals}). The values in the legend correspond to the creep parameters. }
	\label{fig:creep_mesaurements}
\end{figure}

As discussed in \ref{Sec:Theoretical_background} and \ref{subsec:Extenden_creep_model}, the presence of the additional Laplace pressure term (see Eq.\eqref{eq:v_Rb}) can lead to a significant change in the velocity of the DW when the radius of the bubble domain is small enough. To validate this theoretical prediction and its implications, we have conducted a series of DW velocity measurements both in the presence and absence of IP fields at fixed $H_z$, while varying the initial radius $R_b$ of the bubble domain.
\label{subsec:Samples}
We investigate two common magnetic Pt/Co multilayer compositions that are known to display sizable DMI \cite{MAG2022}. Sample $a_4$ has a composition of Ta(5) / Pt(3) / Co(0.8) / Ir(1) / Ta(3) , while Sample $a_3$ has composition  Ta(5) / Pt(3) / Co(0.8) / Ir(3) / Ta(3) where the bracketed numbers refer to the thickness in nm (see TABLE \ref{tab:sample_comp}.).\\ The saturation magnetization $M_s$ was measured using a SQUID-VSM (vibrating sample magnetometer). The effective anisotropy $K_{\text{eff}}$ in the thin film limit can be approximated by \cite{Aharoni1998}
\begin{equation}
	K_{\text{eff}} = K_u - \frac{1}{2}\mu_0 M_s^2,
\end{equation}
where $K_u$ represents uniaxial magnetocrystalline anisotropy and $ -1/2\mu_0 M_s^2$ represents the demagnetization energy density contribution and is obtained by measuring magneto-optical Kerr rotation loops as a function of an in-plane magnetic field. The results are then fitted by minimizing the energy density $E = K_{\text{eff}} \sin^2 \theta + K_{\text{eff}} \sin^4 \theta - H_{ext} M_s \cos(\theta - \phi)$, where $\theta$ is the angle between the applied field $H_{ext}$ and $M_s$, and $\phi$ is the angle between $H$ and the easy axis \cite{HerreraDiez2015}. \\
The exchange stiffness was extrapolated by measuring the velocity as a function of initial bubble radius, following model predictions (see Section \ref{sec:model-predictions} and \ref{Sec:Results} for the details). The error could not always be estimated, as fitting on sample $a_3$ resulted in numerical failure, requiring a manual selection of $A$. All measured values are reported in TABLE \ref{tab:sample_vals}.
The extraction of the creep parameters $\alpha_0, v_0$ necessary for the correct estimation of the DW velocity in the model of eq.\eqref{eq:v_Rb}, is performed on the velocity curve of the DW as a function of the applied field $H_z$ in the creep regime. Eq.\eqref{eq:v_Hz_standard} in the absence of IP fields (i.e. $\mu_0 H_x = 0$) can be expressed as \cite{LEM1998}
\begin{align}
\ln{v} = \ln{v_0} - \alpha_0(\mu_0 H_z)^{-1/4}, 
\label{eq:creep_law_fits}
\end{align}
which can then be used as a linear fit on the experimental data to extract $\alpha_0$ (corresponding to the slope) and $ v_0$ (corresponding to the intercept) as can be seen in Fig.\ref{fig:creep_mesaurements}. To obtain an independent estimate of the value of the DMI coupling constant, $D$, sample $a_3$
has been studied also by Brillouin light scattering (BLS). As a matter of fact, by measuring
the Stokes/Anti-Stokes peak asymmetry it is possible to extract a value from $D$, as
explained in detail in ref.\cite{MAG2022}. BLS measurements have been repeated in several different
areas of the sample surface in order to obtain an average value which could be
representative of the whole sample. The resulting value is $D_{BLS}(a_4)$ = $-0.8 \pm 0.1$ mJ/m$^2$.

\begin{table}
\centering
\begin{tabular}{|c|c|c|c|}
\hline
Sample & Bottom layer (nm) & FM layer (nm) &  Top layer (nm)\\ \hline
$a_4$ & Ta(5)/Pt(3) & Co (0.8)  &    Ir(1)/Ta(3) \\ \hline
 $a_3$&Ta(5)/Pt(3) &  Co (0.8)& Ir(3)/Ta(3)\\\hline
\end{tabular}
\caption{Co-Based samples with Pt and Ir heavy metal (HM) layers}
\label{tab:sample_comp}
\end{table}

\begin{table*}

\centering
\begin{tabular}{|c|c|c|c|c|c|}
\hline
Sample & $M_s$ [MA/m] & $K_{\text{eff}}$ [MJ/m$^3$] & A [pJ/m] & $\alpha_0$ [T$^{1/4}$] & $\ln{v_0}$ [-] \\ 
\hline
$a_4$ & $1.20 \pm 0.09$ & $0.513 \pm 0.002$ & $11 \pm 2$   & $15.5 \pm 0.4$ & $40 \pm 1$ \\ 
\hline
$a_3$ & $1.14 \pm 0.05$ & $0.414 \pm 0.001$ & $10$ &  $20 \pm 1$ & $49 \pm 3$ \\ 
\hline
\end{tabular}
\caption{Experimentally obtained physical parameters of the $a_4$ and n$_{18}$ samples. $M_s$ represents the saturation magnetization, $K_{\text{eff}}$ represents the effective anisotropy, $A$ represents the exchange stiffness. $\alpha_0$ and $\ln{v_0}$ represent respectively, the slope and the intercept of the creep data (see \ref{subsec:Samples} and Fig.\ref{fig:creep_mesaurements}). }
\label{tab:sample_vals}
\end{table*}
\subsection{Results and discussion}
\label{Sec:Results}

\begin{figure}[h!]
	\centering
	\includegraphics[width=\linewidth]{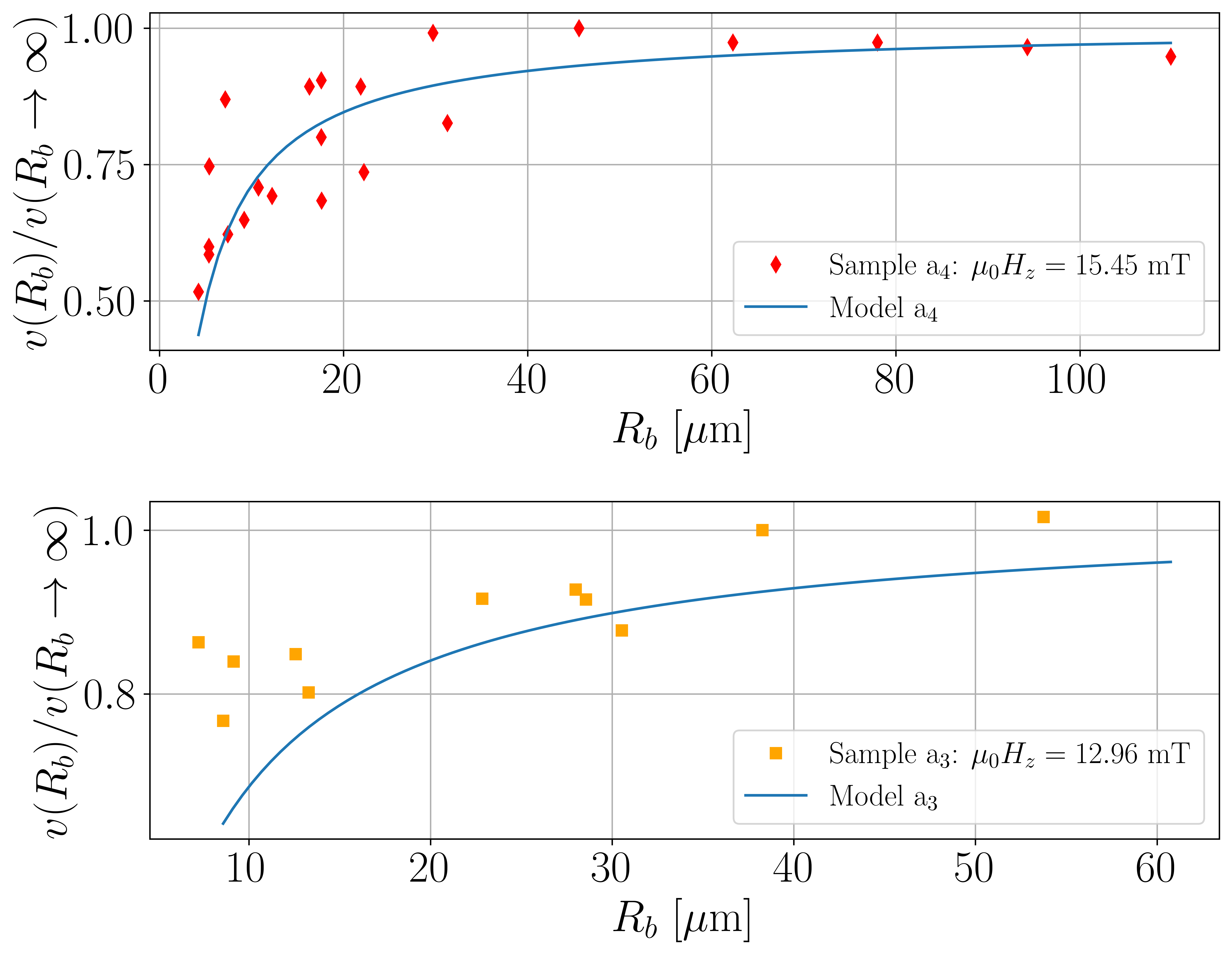}
	\caption{Top: Radius dependence of the DW velocity in the $a_4$ sample (see TABLE \ref{tab:sample_vals}). The blue curve represents a best fit of exchange stiffness $A$ performed on the function of Eq.\eqref{eq:v_Rb}. The parameters used are $M_s = 1.2 $ MA/m , $K_{\text{eff}} = 0.513$ MJ/m$^3$, $\alpha_0 = 15.47$ T$^{1/4}$, $\ln v_0 = 40.5$.  Bottom: Radius dependence of the DW velocity in the $a_3$
		sample (see TABLE \ref{tab:sample_vals}). The blue curve represents a suitable choice of exchange stiffness $A$ performed on the function of Eq.\eqref{eq:v_Rb}. In this case a numerical failure prevented a true fitting procedure, and thus, an error estimation. The parameters used are $M_s = 1.14$ MA/m , $K_{\text{eff}} = 0.414 $ MJ/m$^3$, $\alpha_0 = 19.62$ T$^{1/4}$, $\ln v_0 = 50.29$ For both curves, a value of $\mu_0 H_D = -20$ mT was used.}
	\label{Fig:Radius_dependence_Fit}
\end{figure}

\begin{figure*}
    \centering
    \includegraphics[width=0.75\linewidth]{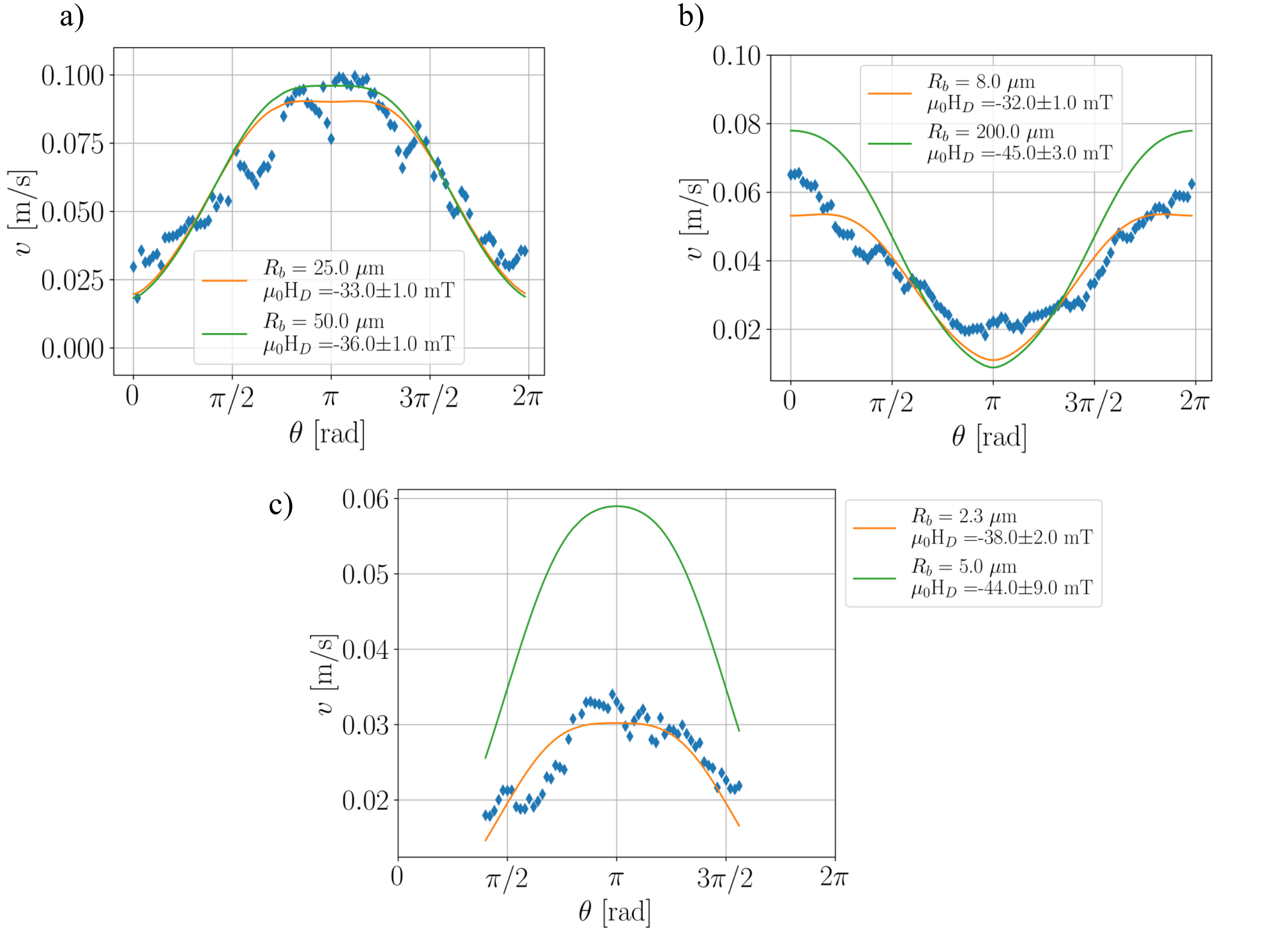}
    \caption{Sample $a_4$: DW velocity as a function of the $\theta$ angle (see Fig.\ref{fig:Setup_bubble}). We show the comparison of the fit between using the experimentally measured $R_b$ (orange curve) and a larger value of $R_b$ (green curve). The $3$ plots represent measurements and fits performed on bubble domains of initial radius $R_b \approx 25 \mu$m and IP field $\mu_0 H_x = 80$ mT (a), $R_b \approx 8 \mu$m and IP field $\mu_0 H_x = -102$ mT (b) $R_b \approx 2.3 \mu$m and IP field $\mu_0 H_x = 110$ mT (c). The parameters used are $M_s = 1.2 $ MA/m , $K_{\text{eff}} = 0.513$ MJ/m$^3$, $\alpha_0 = 15.47$ T$^{1/4}$, $\ln v_0 = 40.5$, $\mu_0 H_z = 13.4$ mT. }
    \label{fig:fits_comparison}
\end{figure*}

\begin{figure}[h]
	\centering
	\begin{minipage}{0.45\textwidth}
		\centering
		\includegraphics[width=\textwidth]{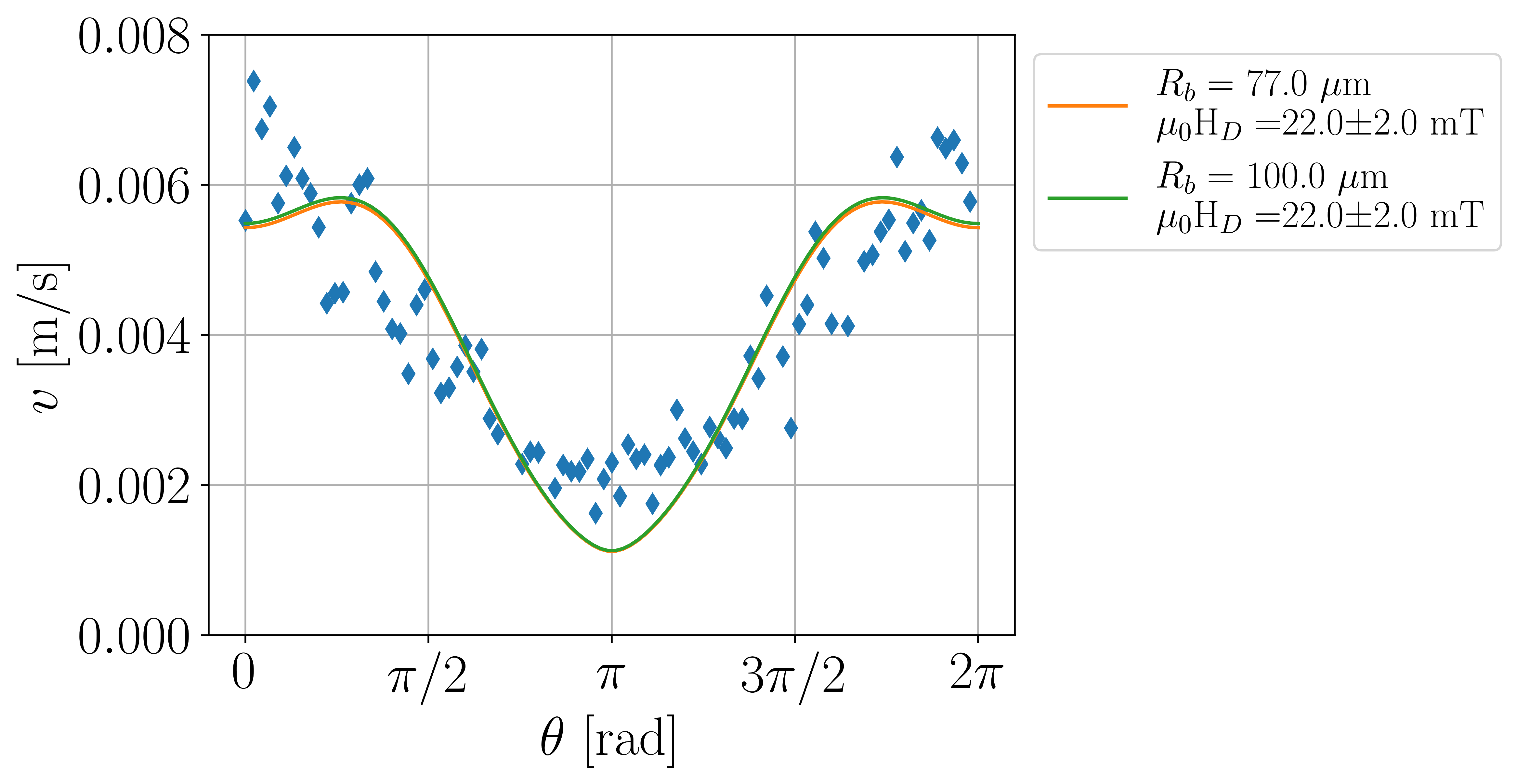}
		\label{fig:plot1}
	\end{minipage}
	\hfill
	\begin{minipage}{0.45\textwidth}
		\centering
		\includegraphics[width=\textwidth]{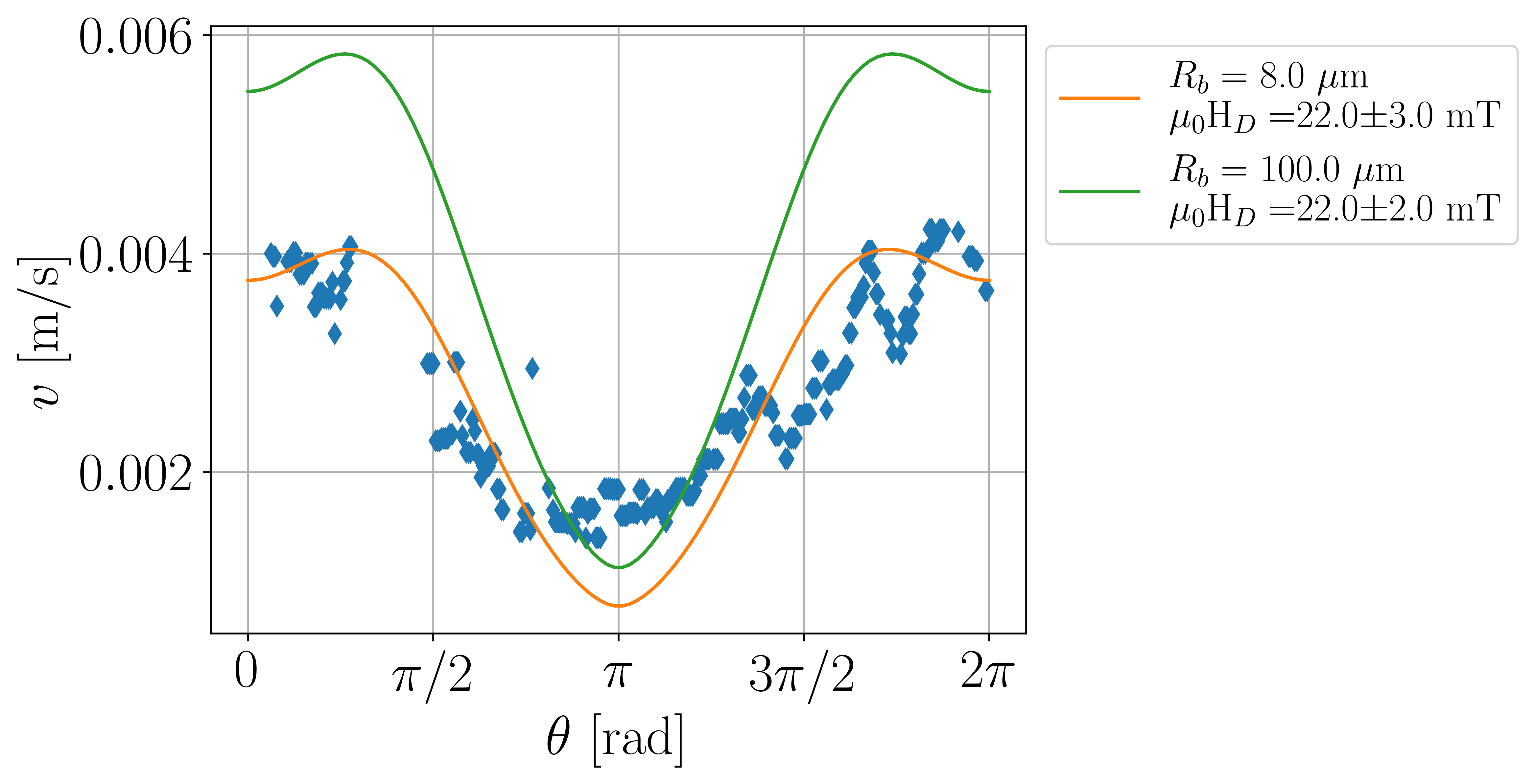}
		\label{fig:plot2}
	\end{minipage}
	\caption{Sample $a_3$: DW velocity as a function of the $\theta$ angle (see Fig.\ref{fig:Setup_bubble}). We show the comparison of the fit between using the experimentally measured $R_b$ (orange curve) and a larger value of $R_b$ (green curve). The $2$ represent measurements and fits performed on bubble domains of initial radius $R_b \approx 77 \mu$m (top), $R_b \approx 8 \mu$m (bottom). Both the measurements were performed with an IP field of $\mu_0 H_x = -80 $ mT. The parameters used are $M_s = 1.14$ MA/m , $K_{\text{eff}} = 0.414 $ MJ/m$^3$, $\alpha_0 = 19.62$ T$^{1/4}$, $\ln v_0 = 50.29$, $\mu_0 H_z = 12.96$ mT.}
	\label{fig:fits_comparison_n9b}
\end{figure}

Fig.\ref{Fig:Radius_dependence_Fit} shows the normalized DW velocity as a function of the initial bubble radius $R_b$. Both samples $a_4$ and $a_3$ display a decrease of the DW velocity below certain values of $R_b$. We furthermore notice how the experimentally measured trends are followed by the theoretical prediction expressed with the velocity formula of Eq.\eqref{eq:v_Rb} (observe the continuous curves in Fig.\ref{Fig:Radius_dependence_Fit}). We fit Eq.\eqref{eq:modified_velocity_Rb} and extract the exchange stiffness $A$ from the data reported in Fig.\ref{Fig:Radius_dependence_Fit}. For both curves in Fig.\ref{Fig:Radius_dependence_Fit}, a value of $\mu_0 H_{D}= -20 $ mT was used. This is reasonable since the DMI has a small influence on $\tilde{R}_b$ with respect to the exchange stiffness in these samples. The obtained values for $A$ are reported in TABLE \ref{tab:sample_vals} and are used in the following. We highlight the fact that the so obtained exchange stiffness values are higher than those obtained in \cite{MAG2022} on the same samples but are closer to literature values reported on similar samples \cite{Hartmann2019}. 
The  radius dependence of the DW of velocity displayed in Fig.\ref{Fig:Radius_dependence_Fit} shows how additional care might be required in the determination of physical quantities measured via DW expansion in the creep regime.\\ The DW Laplace pressure may have a strong influence on the experimental determination of DW speed from bubble expansion experiments in the creep regime. This will necessarily lead to implications for measurement purposes as its the case for the DMI strength measured via the protocol proposed by \cite{Pakam2024} (see \ref{subsec:Bubble_expansions_method}).  We therefore perform domain expansion experiments under the application of IP fields on a variety of bubble sizes. In particular, we measure asymmetric bubble expansion with an IP field for bubbles with an initial radius of $R_b \approx 2.3 \, \mu$m,  8 $\mu$m ,  25 $\mu$m sample $a_4$ (see Fig.\ref{fig:fits_comparison})  and $R_b\approx$  8 $\mu$m, 77 $\mu$m and in sample $a_3$ (see Fig.\ref{fig:fits_comparison_n9b}).
After the experimental determination of $v(\theta)$ (see Fig.\ref{fig:DW_equilibrium_and_velocity}) for the different bubbles, we proceed and fit $v(\theta, R_b)$ from Eq.\eqref{eq:v_Rb} on the measured data in the samples $a_4$ and $a_3$. To highlight the initial radius $R_b$ dependence, we use the following scheme: in one case we use a larger $R_b$ value in the $v(\theta, R_b)$ from Eq.\eqref{eq:v_Rb} to show the effect of neglecting the correct bubble radius (see the green curves in Figs.\ref{fig:fits_comparison} and \ref{fig:fits_comparison_n9b}), while in the other case we use the experimentally measured bubble radius $R_b$ (see the orange curves in Figs.\ref{fig:fits_comparison} and \ref{fig:fits_comparison_n9b}). 

Focusing on sample $a_4$, we can observe (Fig.\ref{fig:fits_comparison}-a) how in the case of large bubbles ($R_b \approx 25 \mu$m), considering the correct radius has a small effect on the success of the fit: fitting the curve with either $R_b = 25 \, \mu$m or $R_b = 60 \,\mu$m causes the value of $\mu_0 H_{D}$ to change by 2 mT. On the other hand, in the regime where the Laplace pressure becomes sizable (see Fig.\ref{fig:Rb_effect_on_V}), the experimentally determined DW velocity becomes highly sensitive to even small variations of $R_b$. The sizable effect of the Laplace pressure needs to be taken in consideration here, as the fits of Fig.\ref{fig:fits_comparison}-b,c (observe the green curves) evidence how the incorrect assumption of the bubble radius $R_b$ result in the failure of the fitting procedure. Using of the correct values of $R_b$ on the other hand  (see the orange curves of Fig.\ref{fig:fits_comparison}-b,c),  shows the success of the fit and how the consistency of the $\mu_0 H_{D}$ is preserved across different bubble dimensions. An analogy can be found using the same procedure on sample $a_3$, where the in Fig.\ref{fig:fits_comparison_n9b}-a), the relatively large initial radius of the bubble (77 $\mu $m) causes the error to be small as compared to Fig.\ref{fig:fits_comparison_n9b}-b), where keeping into account the correct value of the initial radius ($R_b \approx 8 \mu $m) has a visible impact on the success of the fit. As a final remark, we emphasize how the so obtained values of the DMI effective field ($\mu_0 H_{D}(a_3) \approx -22.1 \pm 0.6$ mT and  $\mu_0 H_{D}(a_4) \approx -33 \pm 4 $  mT ) are in agreement with the values reported in \cite{MAG2022}. The higher values of the DMI energy obtained through $D = \mu_0 H_{D} M_s \Delta$  ($D(a_3) = -0.124 \pm 0.006$ mJ/m$^2$ ,  $D(a_4) = -0.18 \pm 0.03$ mJ/m$^2$) are to be related to the higher values of exchange stiffness $A$. We also note that the higher $D$ values are closer (yet still smaller) than DMI energy strength values obtained through independent Brillouin light scattering (BLS) measurements ($D_{BLS}(a_4) = -0.8 \pm 0.1$ mJ/m$^2$ and $D_{BLS}(a_3) = -0.5 \pm 0.1$ mJ/m$^2$ \cite{Kuepferling2023}). The discrepancy between DMI values reported via DW speed measurements in the creep regime and BLS measurements is a well known issue \cite{Kuepferling2023} that will be addressed in future work. 
The modified creep model of Eq.\eqref{eq:v_Rb}, which allows to account for the additional Laplace pressure felt by magnetic bubble domains, eliminates the inconsistency of the DMI values obtained by fitting the $v(\theta)$ curves with the $R_b \rightarrow \infty$ model (i.e. that of Eq.\eqref{eq:v_Hz_standard}). This does not only improve the measurement reproducibility, but can prove necessary in cases where the actual space to allow for a complete expansion of a bubble domain under IP field without impinging on sample barriers or other bubbles is lacking. In particular, the method discussed in \cite{pakam_anisotropic_2024} enhanced by the radius dependence of the velocity of Eq.\eqref{eq:v_Rb} could allow for the in-situ measurements of DMI on patterned samples for device applications in which the surface area of magnetic material might be severely reduced as compared to a full homogeneous films.

\section{Conclusions}\label{sec:conclusions} 

In this paper, we demonstrate how the DW velocity of magnetic bubble domains in the creep regime displays a significant reduction below a sample-dependent threshold radius \cite{MOO2011} in thin film Pt(3)/Co(0.8)/Ir(1) and Pt(3)/Co(0.8)/Ir(3) samples. We show that this phenomenon is due to the Laplace pressure felt by the magnetic bubble and provide an extension of the creep model for DW motion able to describe this dependence. To have a consistent description of the phenomenon, we have to include a radius dependent term both in the energy density of the DW, and in the driving force term of the creep law (see Eq. \eqref{eq:DW_Rb}). To highlight the possible consequences of neglecting this term in models describing DW motion in the creep regime, we study the case of the determination of the DMI strength from asymmetric bubble expansion \cite{Pakam2024,JE2013}.
We show how accounting for the Laplace pressure term is necessary for obtaining reproducible DMI measurements when dealing with bubbles of varying size (Figs.\ref{fig:fits_comparison},\ref{fig:fits_comparison_n9b}).We also stress the fact that the radius dependence becomes increasingly relevant when the creep regime is only accessible with low OOP fields.
Therefore, we conclude that a reproducible measurement of magnetic properties related to creep DW velocity in magnetic bubbles requires an initial assessment and study of the radius dependence of DW velocity. The technique introduced in ref.\cite{Pakam2024}, when coupled with the radius dependent correction of Eq.\eqref{eq:v_Rb}, can be used to measure the DMI on reduced portions of magnetic materials such as patterned surfaces, as might be the case for in-situ measurements of finished devices.

\section{Acknowledgment}
We thank Stefania Pizzini and Laurent Ranno for the fruitful and inspiring discussions. This project is supported by Italian ministry of education PRIN 2022 "Metrology for spintronics: A machine learning approach for the reliable determination of the Dzyaloshinskii-Moriya interaction (MetroSpin)", Grant no. 2022SAYARY.

\appendix
\section{Derivation of domain wall energy density for a bubble domain }
\label{Appendix A:Derivation of domain wall energy density for a bubble domain}
In the following we provide a brief reminder of how to derive the domain wall energy density for a curved domain as opposed to the usual derivation for linear domains \cite{ThiavilleBook2006}. The starting point is the energy density of a magnetized body of volume $V$ with PMA 
\begin{align}
 E = \displaystyle \int_V \bigg\{\underbrace{A|\nabla\bm m|^2}_{\mathcal E_{exch}} - \underbrace{D[m_z(\nabla \cdot \bm m) - (\nabla \cdot \bm m)m_z]}_{\mathcal E_{DMI}} \nonumber
 \\ - \underbrace{ \frac{1}{2}\mu_0 M_s \bm m \cdot \bm H_d  + K_u (\bm m \cdot \bm{\hat u_z})^2 }_{\mathcal E_{Anis}} - \underbrace{\mu_0 M_s \bm m \cdot \bm H}_{\mathcal E_{Zeeman}}  \bigg \} d^3 \bm r, \label{eq:energy_integral}
\end{align}
where $\bm m(\bm r, t) = \bm M(\bm r, t)/M_s $ is the normalized magnetization vector,  $A$ represents the exchange stiffness, $ $$D$ represents the DMI constant, $\bm H_d$ is the magnetostatic field and $\bm H$ is the applied field. If we express the normalized magnetization vector in spherical coordinates $\bm m(\bm r , t) = (\sin \tilde{\theta} \cos \phi, \sin \tilde{\theta} \sin \phi, \cos \tilde{\theta})^T$ (where the angles $\tilde{\theta}(\bm r , t), \phi (\bm r , t) $ manifestly depend on the position $\bm r$ and the time instant $t$), we can write the different energy terms as
\begin{align}
\mathcal E_{exch} &= A[(\nabla\tilde{\theta})^2 + \sin^2 \tilde{\theta} (\nabla \phi)^2] ,\\  \mathcal{E}_{\text{DMI}} &= D \bigg[ 
\cos \phi \frac{\partial \tilde{\theta}}{\partial x} 
+ \sin \phi \frac{\partial \tilde{\theta}}{\partial y} \nonumber
\\ &+ \sin \tilde{\theta} \cos \tilde{\theta} \bigg( 
\sin \phi \frac{\partial \phi}{\partial x} 
- \cos \phi \frac{\partial \phi}{\partial y} 
\bigg) \bigg] ,\\ \mathcal E_{Anis} &= K_0\sin^2 \tilde{\theta} + K \sin^2 \tilde{\theta}\sin^2 \phi ,\\ \mathcal E_{Zeeman} &= \mu_0 M_s(H_x \sin \tilde{\theta} \cos \phi + H_z \cos \tilde{\theta}).
\end{align}
We have written the $\mathcal E_{Anis}$  term in a compact fashion as is usually done in the literature \cite{ThiavilleBook2006}: $K_0$ represents the effective anisotropy, while $K$ is the shape anisotropy. Converting the integration to cylindrical coordinates
\begin{align}
x \rightarrow r \cos \theta \\y \rightarrow r \sin \theta \\ z \rightarrow z^\prime 
\end{align}
and making use of the Bloch Ansatz for the DW profile \cite{ThiavilleBook2006}
\begin{align}
	\begin{cases}\tan(\tilde{\theta}(r)/2) = \exp\bigg(\frac{r - R_b}{\Delta}\bigg) \\ \phi(\theta ) = \theta + \Omega \end{cases} \label{eq:Bloch_Ansatz}
\end{align}
we can rewrite the integral of Eq.\eqref{eq:energy_integral} as
\begin{align}
E &= \delta \int_0^{2\pi} \int_0^{R_t} \bigg\{ A \bigg[ \left( \frac{\partial \tilde{\theta}}{\partial r} \right)^2 + \frac{\sin^2 \tilde{\theta}}{r^2} \bigg] - \nonumber \\ & D \bigg[ \frac{\partial \tilde{\theta}}{\partial r} + \frac{\cos \tilde{\theta} \sin \tilde{\theta}}{r} \bigg] \cos \Omega + \nonumber \\ &\big(K_0 + K \sin^2 \Omega\big) \sin^2 \tilde{\theta} - \mu_0 M_s H_z \cos \tilde{\theta} \bigg\} r \, dr \, d\tilde{\theta},
\end{align}
where $\delta$ represents the thickness of the sample. To obtain the above form, we have exploited the fact that the energy density is independent of the z coordinate, i.e. the energy density is assumed constant along the $ z$ direction. At this point, to derive the energy density of the DW, we have to integrate out the $r$ variable. To do so, we first of all notice that, from the Ansatz of Eq.\eqref{eq:Bloch_Ansatz} we have $\partial \tilde{\theta}/\partial r = \sin \tilde{\theta}/\Delta = \frac{1}{\Delta}\text{sech}\big(\frac{r - R_b}{\Delta}\big)$.  Operating the substitution $\eta = \frac{r - R_b}{\Delta}$ we obtain
\begin{align}
	E &= \delta \int_0^{2\pi} \int_{-R_b/\Delta}^{(R_t - R_b)/\Delta} \bigg\{ A \sin^2 \tilde{\theta} \bigg[  \frac{1}{\Delta^2}  + \frac{1 }{(\Delta\eta + R_b)^2} \bigg] - \nonumber \\
	 &D \bigg[ \frac{\sin \tilde{\theta}}{\Delta} + \frac{\cos \tilde{\theta} \sin \tilde{\theta}}{\Delta\eta + R_b} \bigg] \cos \Omega + \big(K_0 + K \sin^2 \Omega\big) \sin^2 \tilde{\theta} \nonumber \\
	  &- \mu_0 M_s(H_z \cos  \tilde{\theta} + H_x \sin \tilde{\theta} \cos \Omega)\bigg\} (\Delta\eta + R_b) \, dr \, d\tilde{\theta},
\end{align}
Which, in the limit $R_b/\Delta \gg 1$ and $\lim_{R_t \rightarrow \infty}$ is asymptotically equivalent to  
\begin{align}
	E &\sim \delta \Delta\int_0^{2\pi} \int_{-\infty}^{\infty} \bigg\{ A \, \text{sech}^2(\eta)\bigg[  \frac{1}{\Delta^2}  + \frac{1 }{(\Delta\eta + R_b)^2} \bigg] - \nonumber \\
	& D \bigg[ \frac{\text{sech}(\eta)}{\Delta} + \frac{\sqrt{1 - \text{sech}^2(\eta)} \, \text{sech}(\eta)}{\Delta\eta + R_b} \bigg] \cos \Omega + \nonumber \\
	& \big(K_0 + K \sin^2 \Omega\big) \text{sech}^2(\eta) - \nonumber \\
	&\mu_0 M_s (H_z \sqrt{1 - \text{sech}^2(\eta)} + H_x \text{sech}(\eta) \cos \Omega  \bigg\} \nonumber \\  &\times (\Delta\eta + R_b) \, d\eta \, d\tilde{\theta}, \label{eq:Integrate_Edens}
\end{align}
We can now use some know integration formulas for hyperbolic functions (see e.g. )\cite{gradstejn_table_2009})
\begin{align}
	&\int_{-\infty}^{\infty} \eta \, \text{sech}^2(\eta) d \eta = 0 \\ &\int_{-\infty}^{\infty}  \, \text{sech}^2(\eta) d \eta = 2 \\ &\int_{-\infty}^{\infty}  \, \text{sech}(\eta) d \eta = \pi \\ &\int_{-\infty}^{\infty}\sqrt{1 - \text{sech}^2(\eta)} \, \text{sech}(\eta) d \eta = 2
\end{align}
As well as the fact that the integrals of the form 
\begin{align}
	\int_{-\infty}^{\infty}  \, \frac{ \text{sech}^2(\eta)}{\Delta \eta + R_b} d \eta,
\end{align}
only converge in the limit $R_b/\Delta \gg 1$, for which we then obtain 
\begin{align}
\int_{-\infty}^{\infty}  \, \frac{ \text{sech}^2(\eta)}{\Delta \eta + R_b} d \eta \approx \frac{2}{\Delta R_b}.
\end{align}
Combing the above mentioned identities with Eq.\eqref{eq:Integrate_Edens} we finally obtain the surface tension of the bubble $\sigma_{DW}$ as a function of the bubble radius $R_b$ and the angle $\Omega$ (see Fig.\ref{fig:Setup_bubble})
\begin{align}
	\frac{E}{2 \pi t R_b} &:= \sigma_{DW} =  \left( 4\sqrt{A K_{\text{eff}}} - \pi D \cos \Omega + 2\Delta K \sin^2 \Omega \right) + \nonumber \\ &\frac{2 A \Delta}{R_b^2} - 2\pi \mu_0 M_s H_z R_b
\end{align}

\section{Brief reminder of creep theory }
\label{Appendix B: : Brief reminder of creep theory}
In the following we provide a brief reminder of the origin of the creep relation reported in Eq.\eqref{eq:creep_law_fits}. For an in depth review of the topic, we suggest \cite{Metaxas2007} and references therein. As mentioned in \ref{Sec:Theoretical_background}, the velocity of a domain wall in the creep regime is determined by the competition of thermal energy and energy barrier. The energy barrier encodes all the energy costs (elastic energy) and gains (Zeeman and pinning energy) the magnetic DW is subject to as a function of its dimension and configuration. The energy barrier is given by Eq.\eqref{eq:Energy_barrier} in the main text. The velocity of the DW is determined by the ability of thermal fluctuations to overcome the highest possible value of the energy barrier, i.e. the extremal value of $ \Delta E(u,L)$ (see Eq.\eqref{eq:Energy_barrier}), where $u$ and $L$ are the geometrical parameters of the DW displacements displayed in Fig.. An important point however, is that $u$ and and $L$ are not independent from one another and are related by the exponential relation
\begin{align}
   u(L) = u_c \bigg(\frac{L}{L_c} \bigg)^\zeta
\end{align}
where $u_c$ represents the transverse scaling parameter \cite{Kardar1985} and $L_c$ represents the so called Larkin length \cite{larkin1979pinning}. The Larkin length $L_c$ is a characteristic length scale obtained by the optimal balance of the pinning energy and the elastic energy of Eq.\eqref{eq:Energy_barrier} when the DW has jumped exactly one pinning center, i.e. $u = \xi$ (perhaps add a small cartoon). 
\begin{align}
    L_c = 2 \bigg( 2 \frac{\bar{\mathcal{E}}_{DW} \xi^2}{\gamma} \bigg)^{1/3}. 
\end{align}
On the other hand, $\zeta$ represents the critical exponent of this relation and is often time referred to as the roughness exponent. Much like the creep critical exponent $\mu = 1/4$ introduced in section (add section), $\zeta = 2/3$ depends on very fundamental properties of the theory such as the dimensionality of the interface etc. \cite{Kolton2005}. Equipped with $u(L)$, we can readily plug this expression in $\Delta E(u(L), L) $ and maximize it to obtain the energy barrier $F_b$.
\begin{align}
    F_b = \max_L \Delta E(u(L),L) = \frac{4}{5^{5/4}} \frac{u_c^{9/4} \sqrt{\gamma}}{\xi} \bigg( \frac{\bar{\mathcal{E}}_{DW}}{H_z M_s}\bigg)^{1/4}
\end{align}
which can then finally be plugged in the velocity to obtain the well known creep law \cite{LEM1998}

$$
v = v_0 \exp(- \alpha_0 H_z^{-1/4}),
$$

where we define $\alpha_0$ as 
\begin{align}
    \alpha_0 := \frac{4}{5^{5/4}} \frac{u_c^{9/4} \sqrt{\gamma}}{k_B T \xi} \bigg( \frac{\bar{\mathcal{E}}_{DW}}{M_s}\bigg)^{1/4} 
\end{align}

\section{Bubble expansion measurement method}
\label{subsec:Bubble_expansions_method}
\begin{figure}
	\centering
	\includegraphics[width=1\linewidth]{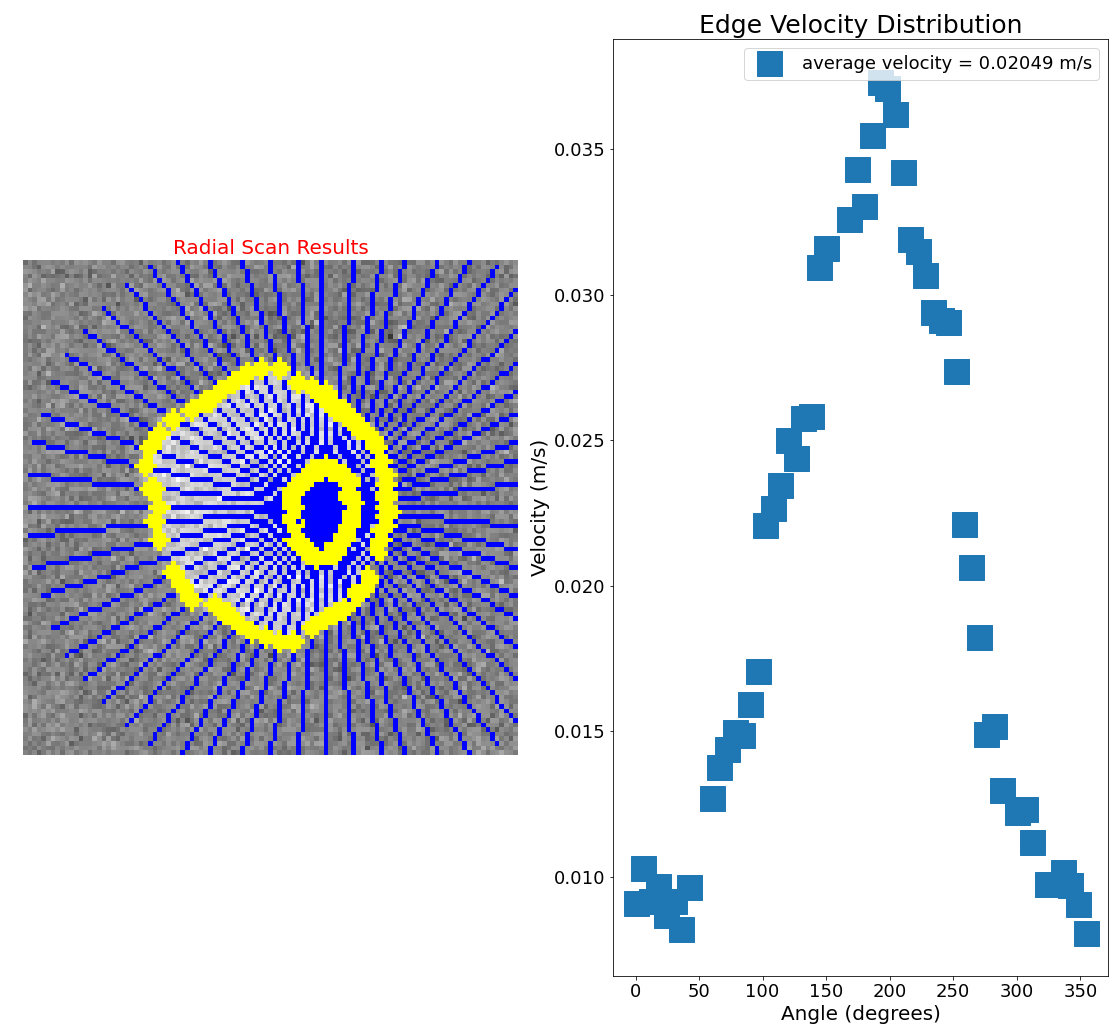}
	\caption{Output of the bubble edge detection code based on the C2V python library \cite{culjak2012brief}.}
	\label{fig:Bubble_edge_detector}
\end{figure}

The analysis of the bubble domain expansion is performed using polar Kerr magneto-optic principles, while the data analysis and the extraction of the velocities as a function of the angle $v(\theta)$ is performed using a custom written python-based user interface code. The user has to provide the center of the initial bubble manually and then the code proceeds to detect the edges using the python C2V library \cite{culjak2012brief}.  Observing  Fig.\ref{fig:Bubble_edge_detector}, the left image represents the result of the edge detection analysis while the right image reports the measured velocities as a function of the $\theta$ angle (see Fig.\ref{fig:DW_equilibrium_and_velocity} in the main text). The so obtained velocity curves as a function of the $\theta$ angle are then used to perform the fits described in the main text (see Figs.\ref{fig:fits_comparison} and \ref{fig:fits_comparison_n9b} in the main text). 
\newpage
\bibliographystyle{apsrev4-2}

%

\end{document}